\documentclass[11pt,a4paper]{article}
\usepackage{amsmath, amsthm} 
\usepackage{amsfonts}
\usepackage{amssymb,graphics,psfrag}
\usepackage[bottom]{footmisc}
\usepackage{hyperref}
\usepackage{slashed}

\def\hybrid{\topmargin -10pt    \oddsidemargin 0pt
        \headheight 0pt \headsep 0pt
       \textwidth 6.25in       
      \textheight 9.5in       
        \marginparwidth .875in
        \parskip 5pt plus 1pt   \jot = 1.5ex}
\hybrid
\numberwithin{equation}{section}
\numberwithin{table}{section}
\begin{document}

\thispagestyle{empty}

\rightline{\small MIFPA-12-03}

\vskip 3cm
\noindent
{\LARGE \bf Boundary Terms, Spinors and Kerr/CFT   }\\ 
\vskip .8cm
\begin{center}
\linethickness{.06cm}
\line(1,0){447}
\end{center}
\vskip .8cm
\noindent
{\large \bf Melanie Becker, Waldemar Schulgin}

\vskip 0.2cm

\hskip -0.1cm  {\em    George and Cynthia Mitchell Institute}
\vskip -0.15cm
{\em \hskip -.05cm for Fundamental Physics and Astronomy}
\vskip -0.15cm
{\em \hskip -.05cm Texas A \&M University, College Station}
\vskip -0.15cm
{\em \hskip -.05cm TX 77843--4242, USA}
\vskip -0.15cm
\vskip 0.5cm
{\tt \hskip -.5cm mbecker AT physics.tamu.edu, schulgin AT physics.tamu.edu}

\vskip 1cm

\vskip0.6cm

\noindent
{\sc Abstract:} 
Similarly as in AdS/CFT, the requirement that the action for spinors 
be stationary for solutions to the Dirac equation with fixed boundary conditions 
determines the form of the boundary term that needs to be added to the standard Dirac action in Kerr/CFT.  
We determine this boundary term and make use of it to calculate the two-point function for spinor fields in 
Kerr/CFT. This two-point function agrees with the correlator of a two dimensional relativistic conformal field theory.

\pagebreak

{\tableofcontents }

\newpage


\section{Introduction}

AdS/CFT and its generalizations play a major role in recent developments of theoretical physics. 
Examples which are related to realistic physical objects are, however, still rare.
The ultimate goal of the Kerr/CFT correspondence is to describe the black holes of our universe
in terms of a dual two dimensional conformal field theory. The concrete proposal of \cite{Guica:2008mu}
is that the near horizon region of a near-extremal Kerr black hole (the so called near-NHEK geometry) is 
dual to a two dimensional conformal field theory.
Even though we are still far from describing a real black hole, many tests supporting this conjecture have appeared in the literature so far (see \cite{Bredberg:2011hp} for a review). In particular, the scattering amplitudes
for spinor fields computed in \cite{Hartman:2009nz} (see also \cite{Chen:2010ni})
were found to be in agreement with the conformal field theory result.
Spinor fields in AdS/CFT and its non-relativistic generalizations are particularly 
interesting to study, as their correlation functions are many times related to semi-realistic physical observables such as the spectral function. 
 
In this note we would like to revisit spinor fields in Kerr/CFT. We would like to consider spinor fields in the near-NHEK geometry
\begin{equation}\label{nearNHEK}
ds^2=2J\Gamma\Big(-r(r+4\pi T_R)dt^2+\frac{dr^2}{r(r+4\pi T_R)}+d\theta^2+\Lambda^2\left(d\phi+(r+2\pi T_R)dt\right)^2\Big) \ ,
\end{equation}
where
\begin{equation*}
\Gamma(\theta)=\frac{1+\cos^2\theta}{2} \ , \qquad \Lambda(\theta)=\frac{2\sin\theta}{1+\cos^2\theta} \ , \qquad\phi\sim\phi+2\pi,\ 0\le\theta\le\pi\ .
\end{equation*}
More concretely, we would like
 to calculate two-point correlation functions for spinor fields in this geometry. 

Recall that in AdS/CFT
 spinor field correlation functions are slightly more involved than correlation functions for scalars.
 Let's recapitulate some highlights for spinor fields in AdS/CFT, which will become handy 
 later on (see \cite{Henningson:1998cd}, \cite{Mueck:1998iz}, \cite{Henneaux:1998ch}, \cite{Iqbal:2009fd} for more details).

The key assumption of the correspondence is the equivalence between the partition functions of a CFT in $d$-dimensions and a bulk gravitational theory in $(d+1)$-dimensions
\begin{equation}\label{eins}
\langle \exp\left( \int d^d x\, \left(\bar\chi_0{\cal O}+\bar{\cal O}\chi_0\right)\right)\rangle_{{\rm QFT}}=e^{-S_{\rm{grav}}(\chi_0,\bar \chi_0)}.
\end{equation}
In this formula $\chi_0$ is the asymptotic value of the $(d+1)$-bulk spinor $\psi$
\begin{equation}\label{zwei}
\lim_{r\rightarrow\infty}\, \psi\sim\chi_0 \, ,
\end{equation}
that couples to the conformal field theory operator ${\cal{O}}$.
The above formula tells us that to calculate correlation functions of the CFT operator ${\cal O}$, one needs to evaluate the gravitational action $S_{\rm{grav}}$ for solutions to the Dirac equation with proper boundary conditions.
The gravitational action functional contains a bulk term described by the standard Dirac action that vanishes for solutions to the equations of motion. In addition there is a boundary term \cite{Henningson:1998cd}, \cite{Mueck:1998iz},\cite{Iqbal:2009fd} which is non-vanishing for solutions to the equations of motion.
Correlation functions of the CFT operator $\cal{O}$ are determined by this boundary term.
For example, the two point function of two conformal field theory operators ${\cal O},\bar {\cal {O}}$ is given by the functional derivative of the boundary term $S_{\rm {bdry}}$ 
\begin{equation}\label{2ptfct}
\langle {\cal O}\bar{{\cal} {O}}\rangle = \frac{\delta^2S_{\rm {bdry} }}{  \delta \bar \chi_0\delta {\chi}_0}.
\end{equation}
As nicely shown in \cite{Henneaux:1998ch}, 
the form of the gravitational boundary term is dictated by the variational principle.
More recently it was shown that different boundary terms all satisfying the variational principle
can be added to the bulk action \cite{Laia:2011zn}. Different boundary terms lead to different conformal field theories. 

Having the explicit form of the boundary term, the main challenge is to find solutions to the Dirac equation with proper boundary conditions. The situation here is a bit more involved than for a simple scalar, because bulk and boundary spinors live in different dimensions and thus have different number of components in the minimal representation.
A formula like \eqref{zwei} needs to be interpreted with more care. Since $\psi$ is a $(d+1)$-dimensional spinor it contains twice as many degrees of freedom as $\chi_0$ that lives in $d$-dimensions. Only half of the components of $\psi$ can be fixed by $\chi_0$. The other half is determined in terms of the first by the Dirac equation. Therefore, $\psi$ is decomposed into two eigenstates of a projection operator
\begin{equation}
\psi=\psi_++\psi_-\ , \qquad \psi_\pm=\Gamma_\pm \psi, \quad \rm{with}\quad \Gamma_{\pm}=\frac{1}{2}(1\pm\Gamma^r).
\end{equation}
The explicit form of the projection operator depends on the dimension of the boundary. Details can be found in e.g  \cite{Iqbal:2009fd}. The upshot is that for generic values of the spinor mass $\mu$, the $\psi_+$ spinor is the leading component in the large $r$ expansion. This spinor corresponds to the source that is fixed by the boundary condition and which couples to the conformal field theory operator\footnote{There is a small range of values for the mass $\mu$ in which $\psi_-$ rather than $\psi_+$ is fixed by boundary conditions.}
\begin{equation}\lim_{r\rightarrow \infty} r^{d/2-\mu}\psi_+=\chi_0\, .
\end{equation}
The spinor $\psi_-$ is determined in terms of $\psi_+$ by the Dirac equation and vanishes as it approaches the boundary.
Once the spinors solving the equations of motion with proper boundary conditions are known, the evaluation of the gravitational action functional (more precisely the boundary term) will lead to the CFT correlation functions, as previously mentioned.

Similar in spirit, in this paper we show that a boundary term needs to be added to the Dirac action for spinor fields
in the near-NHEK geometry for the variational principle to be satisfied.
The boundary term is the key ingredient for the calculation of the fermionic correlation functions. Using the proposed boundary term, it is shown that the bulk fermionic two-point function agrees with the two-point function of a two dimensional conformal field theory.
Some additional care, however, is required because we shall perform our calculation in Lorentzian signature, rather than analytically
continuing to Euclidean signature. The reason is that we are not aware of an Euclidean version of the near-NHEK metric.
A Lorentzian version of AdS/CFT (where the action carries an `$i$')

\begin{equation}\label{minpart}
\langle \exp\left( i\int d^d x\, \left(\bar\chi_0{\cal O}+\bar{\cal O}\chi_0\right)\right)\rangle_{{\rm QFT}}=e^{-iS_{\rm{grav}}(\chi_0,\bar \chi_0)},
\end{equation}
leads to some additional subtleties that are well known in the context of AdS/CFT (see \cite{Marolf:2004fy} for a discussion). 
As first explained in \cite{Son:2002sd} (and later reformulated by \cite{Iqbal:2009fd}), having complex solutions to the equations of motion requires us to amend the Lorentzian version of AdS/CFT with some further constraints: 

(1) To evaluate the action functional appearing on the right hand side of \eqref{minpart} we should consider the solutions to the equations of motion with incoming boundary conditions. (2) To evaluate boundary terms of the action, we should not consider any contributions coming from the horizon. (3) Applying the Euclidean AdS/CFT prescription to the Lorentzian theory means that the desired correlator plus its complex conjugate appear once the functional derivative of the gravitational action functional is taken. The correct result for the correlator is given by one of these contributions, while the other should be discarded.  
We shall see that these three constraints plus the equivalence of partition functions \eqref{minpart} provides the correct fermionic correlation function in Kerr/CFT.
This paper is organized as follow. 
In Section 2 we consider the variational principle for spinor fields in Kerr/CFT and determine the boundary term. 
In Section 3 we perform the calculation of the fermionic two-point function using the proposed boundary term.
In Section 4 we show how the result of Section 3 can be matched with a two dimensional relativistic conformal field theory.  Our conclusions appear in Section 5.
In Appendix \ref{appA} we present the features
of the near-NHEK geometry we need for the calculation of correlation functions, while Appendix \ref{appB} is left for notations and conventions.

\section{Variational Principle}
The bulk action for fermions with mass $\mu$ in the near-NHEK geometry is the standard Dirac action
\begin{equation}\label{bulk}
S_{\rm {bulk}}=i\int d^4 x \sqrt{-g}\bar \psi \left( {\slashed D} -\mu\right)\psi,
\end{equation}
where we have dropped an overall normalization factor. 
We use the representation of the four dimensional bulk gamma matrices as given in appendix \ref{appB}. 
To determine the boundary conditions on the spinor we calculate the variation of the action
which is given by \footnote{The boundary of the near-NHEK geometry is described by large but finite $r$,  $r_B\gg 1$, such that $r_+-r_-\ll \lambda r_B\ll 1$, where $\lambda$ goes to zero. $r_+, r_-$ are the positions of the outer and inner horizons of the Kerr black hole.} 

\begin{equation}\label{nine}
\delta S_{\rm{bulk}}=i\int_{r=r_B}\, d^3x \sqrt{-g_B} \bar \psi \Gamma^r \delta \psi+\ldots,
\end{equation}
where the dots denote terms that vanish by the equations of motion. Here $g_B=g g^{rr}$ describes
the induced boundary metric and $r_B$ is the cutoff describing the boundary of the near-NHEK geometry. 
The gamma matrix $\Gamma^r$ in the near-NHEK geometry takes the form 

\begin{equation}
\Gamma^r=-\frac{r(r+4\pi T_R)}{8J\Gamma}(\Gamma^0+\Gamma^3)+\frac{1}{2}(\Gamma^0-\Gamma^3).
\end{equation}
Boundary conditions need to be imposed so that the variation of the gravitational action vanishes.
To do so, we take a closer look at the spinor solving the Dirac equation
\begin{equation}
({\slashed D} -\mu)\psi=0\, .
\end{equation}
The solution to this equation in the Kerr geometry was worked out by Chandrasekhar in the late seventies \cite{Chandrasekhar:1976ap}.
Using the Newman-Penrose formalism he showed that the Dirac equation can be separated into a radial and an angular equation. Finding an analytical expression for the solution proved, nevertheless, to be very difficult. For a long time only numerical solutions were available. More than thirty years later, an analytic expression for the solution to the Dirac equation in the near-NHEK limit 
was obtained in \cite{Hartman:2009nz}. In this limit (described to the necessary details in Appendix \ref{appA}) the spinor computed in \cite{Hartman:2009nz} takes the form
\begin{equation}\label{spinorsolution}
\psi=e^{-i n_R t+i n_L \phi}
\left(\begin{matrix}
-R_{1/2}S_{1/2}\\ \\
\frac{R_{-1/2}S_{-1/2}}{\sqrt{2}M(1-i\cos(\theta))}\\\\
-\frac{R_{-1/2}S_{1/2}}{\sqrt{2}M(1+i\cos(\theta))}\\\\
R_{1/2}S_{-1/2}\\ 
\end{matrix}\right)\, ,
\end{equation}
where $R_{\pm 1/2}= R_{\pm 1/2}(r)$
describes the radial dependence and $S_{\pm 1/2}=S_{\pm 1/2}(\theta)$. 
Even though the radial part $R_{\pm 1/2}$ of the solution is in general a hypergeometric function, we only need its asymptotic expression (for large but finite $\lambda r$.  The solution with infalling boundary conditions is 
\begin{eqnarray}\label{R}
R_{1/2}(r)&=&N_{1/2}T_R^{-in_R/2-1/2}\left(A_{1/2}\left(\frac{ r}{T_R}\right)^{-1+\beta}+B_{1/2}\left(\frac{ r}{T_R}\right)^{-1-\beta}\right)+\ldots\\
R_{-1/2}(r)&=&N_{-1/2}T_R^{-in_R/2+1/2}\left(A_{-1/2}\left(\frac{ r}{T_R}\right)^{\beta}+B_{-1/2}\left(\frac{r}{T_R}\right)^{-\beta}\right)+\ldots
\end{eqnarray}
The coefficients appearing in these expressions are defined in terms of gamma functions
\begin{eqnarray}\label{AB}
A_s&=&\frac{\Gamma(1-i(n_R+n_L)-s)\Gamma(2\beta)}{\Gamma(\frac{1}{2}+\beta-in_R)\Gamma(\frac{1}{2}+\beta-in_L-s)}\ , \nonumber\\
B_s&=&\frac{\Gamma(1-i(n_R+n_L)-s)\Gamma(-2\beta)}{\Gamma(\frac{1}{2}-\beta-in_R)\Gamma(\frac{1}{2}-\beta-in_L-s)} \ .
\end{eqnarray}
The $N$'s describe normalization factors
\begin{equation}\label{beta}
\frac{N_{1/2}}{N_{-1/2}}=\frac{1/2-i(n_R+n_L)}{M(\Lambda_\ell+i\mu M)}\, ,\qquad \beta^2+n_L^2=\Lambda_\ell^2+\mu^2 M^2.
\end{equation}
In this paper we restrict to real values of $\beta$ for simplicity.
Similarly as in AdS/CFT, only half of the components of $\psi$ can be fixed at the boundary
(the other half is related to the first half by the Dirac equation and will vanish at the boundary). 
To decide which components of $\psi$ we would like to fix, 
it is convenient to introduce projection operators
\begin{equation}
P_\pm=\frac{1}{2}\left(1\pm\Gamma^0\Gamma^3\right)\, ,
\end{equation}
which satisfy $P_+^2=P_+$, $P_-^2=P_-$ and 
\begin{equation}
\Gamma^0\pm \Gamma^3=\Gamma^0 P_\pm=P_\mp \Gamma^0.
\end{equation}
These operators allow us to write the bulk spinor in terms of projector eigenstates as 
\begin{equation}
\psi=\psi_++\psi_- \, .
\end{equation}
Here $\psi_+$ satisfies
\begin{equation}
P_+\psi=\psi_+=e^{-in_R t+i n_L\phi}R_{1/2}
\left(\begin{matrix}
-S_{1/2} \\
0\\
0\\
S_{-1/2}\\ 
\end{matrix}\right)\, ,
\end{equation}
while
$\psi_-$ obeys
\begin{equation}
P_-\psi=\psi_-=e^{-in_R t+i n_L\phi}\frac{R_{-1/2}}{\sqrt{2}M}
\left(\begin{matrix}
0\\ \frac{S_{-1/2}}{1-i\cos \theta} \\
-\frac{S_{1/2}}{1+i\cos \theta}\\
0
\end{matrix}\right),
\end{equation}
and conjugate spinors satisfy
\begin{equation}
\bar\psi P_\pm=\bar\psi_\mp.
\end{equation}
To decide whether $\psi_+$ or $\psi_-$ is the source (which gets fixed at the boundary), we notice that there is a relation 
between both boundary spinors\footnote{The precise relation is given in the next section.}
\begin{equation}
\psi^{B}_+\sim { R_{1/2}^B \over  R_{-1/2}^B}\psi^{B}_-,
\end{equation}
where the index $B$ denotes boundary quantities.  Taking into account \eqref{AB}, this relation tells us that for real $\beta$ we should treat $\psi^B_-$
as the source, while $\psi^B_+$  vanishes at the boundary. 

We can now proceed to evaluate the boundary term. To do so it is convenient to write the $\Gamma^r$ matrix in terms of the projection operators
 \begin{equation}\label{ten}
 \Gamma^r=\frac{1}{8J\Gamma}r(r+4\pi T_R)P_-\Gamma^0 P_+-\frac{1}{2}P_+\Gamma^0 P_-.
 \end{equation}
 It is easy to see that the boundary term \eqref{nine} becomes 
 
 \begin{equation}\label{eleven}
 \delta S_{\rm {bulk}}=i\int_{r=r_B}\, d^3x\, \sqrt{-g_B}\left( \frac{1}{8J\Gamma}\, r_B(r_B+4\pi T_R)\, \psi^\dagger_+\delta\psi_+-\frac{1}{2}\psi_-^\dagger\delta\psi_-\right).
 \end{equation}
 We had seen that 
 $\psi_-$ is the source, so this spinor and its conjugate are fixed at the boundary
 \begin{equation}\label{cond}
 \delta\psi_-\Big|_{r_B}=0 \, ,\quad  \delta\bar \psi_-\Big|_{r_B}=0 \, .
 \end{equation}
 To cancel the contribution proportional to $\delta \psi_+$ we need to add a boundary term
\begin{equation}\label{bndry_term}
S_{\rm{bdry}}
=-\frac{r_B(r_B+4\pi T_R)}{8J\Gamma}\, i\int_{r=r_B} d^3 x \, \sqrt{-g_B}\, \psi_+^\dagger\psi_+.
\end{equation}
This guarantees that the variation of the total action vanishes\footnote{Here we used $\delta\psi^\dagger_+\Big|_{r_B}=0$, since $\psi_+^\dagger\sim \bar \psi_-$.}
\begin{equation}\label{twelve}
 \delta S_{\rm {total}}=\delta S_{\rm {bulk}} +  \delta S_{\rm {bdry}}=0.
 \end{equation}
It is interesting to observe that the boundary term \eqref{bndry_term} looks similar to the non-relativistic boundary terms recently considered in \cite{Laia:2011zn}. There it was argued that non-relativistic conformal field theories can be generated through Lorentz violating boundary terms, even though the underlying bulk theory is Lorentz invariant. One may wonder if the conformal field theory dual to the near-NHEK geometry could be non-relativistic.
Some recent discussion on the possible connection between Kerr/CFT and non-relativistic conformal field theory has appeared recently in the literature \cite{ElShowk:2011cm}. A more extensive analysis is needed to answer this question.

It is interesting to notice that the boundary term can be written as
\begin{equation}\label{bound}
S_{\rm {bdry}}=i\int_{r=r_B}\, d^3 x \sqrt{-g_B} \, \bar \psi \Gamma^r\psi,
\end{equation}
up to contact terms. This expression is familiar from the fermionic flux derived in \cite{Martellini:1977qf}, \cite{Iyer:1978du}.
There it was shown that superradiance does not occur for a fermionic field in a Kerr geometry as the particle flux into the black hole is always positive. Precisely the same expression for the fermionic flux entered the scattering calculation done in 
\cite{Hartman:2009nz}, so the above boundary term does not come as a surprise. 
It is nice to see this expression emerge from the variational principle.

\section{Fermionic Two-Point Function: the Bulk}

To calculate correlation functions for spinors in the near-NHEK geometry we need to evaluate the boundary term \eqref{bound} for spinors satisfying the equations of motion.
We would like to express the bulk spinor \eqref{spinorsolution} in terms of its value on the boundary. To do so it is convenient to off the $\theta$ dependence by introducing spinors 
$a^{\pm}$ 

\begin{equation}
\psi=e^{-i n_R t+i n_L \phi}\left( R_{1/2}
\underbrace{\left(\begin{matrix}-S_{1/2}\\0 \\ 0 \\S_{-1/2}
\end{matrix}
\right)}_{a_+}
+R_{-1/2}
\underbrace{\left(
\begin{matrix}
A & 0&0&0\\
0&\frac{1}{\sqrt{2}M(1-i\cos \theta)}&0&0\\
0&0&\frac{1}{\sqrt{2}M(1+i\cos \theta)}&0\\
0 & 0&0&B
\end{matrix}
\right)}_{Z}
\underbrace{\left(\begin{matrix}0\\S_{-1/2}\\-S_{1/2}\\0
\end{matrix}
\right)}_{a_-}
\right)
\end{equation}
$A$ and $B$ are arbitrary non-zero entries, so that $Z$ is invertible.
The eigenstates of the projection operator $\psi_\pm$ can be conveniently written as
\begin{eqnarray}\label{fourfour}
\psi_+&=&e^{-in_R t+in_L \phi}R_{1/2} a_+,\nonumber\\
\psi_-&=&e^{-in_R t+in_L \phi}R_{-1/2}Za_-.
\end{eqnarray}

Since there is a relation between $a_+$ and $a_-$,  $\Gamma^0 a^+=a^-$, we can write the bulk spinor in terms of $a_-$ only 
\begin{equation}
\psi=e^{-in_R t+i n_L \phi}\left( R_{1/2}\Gamma^0+R_{-1/2}Z \right) a_-.
\end{equation}
Using equation (\eqref{fourfour}) we can express this spinor and its conjugate in terms of boundary data
\begin{eqnarray}
\psi&=&\left(R_{1/2}\Gamma^0 Z^{-1}+R_{-1/2}\right)\frac{\psi^B_-}{R_{-1/2}^B},\nonumber\\
\bar \psi& =& \frac{\bar \psi^{B}_-}{\bar R_{-1/2}^B}\left(\bar R_{1/2} \Gamma^0Z^{*-1}+\bar R_{-1/2}\right),
\end{eqnarray}
where the bar on $\bar R_{\pm 1/2}$ means complex conjugation.
To apply the prescription \eqref{2ptfct} for computing the boundary two-point function, we would like to express the boundary term as a double integral over momenta. This will allow us to take the functional derivative. Fourier transforming along the $t$ are $\phi$ directions we introduce new spinors 
\begin{eqnarray}
\psi_F (r,\theta,n_L,n_R)&=&\delta(n_L-n_L')\delta(n_R-n_R')\Big(R_{1/2} (r, n_L',n_R')\Gamma^0 Z^{-1}+R_{-1/2}(n_L',n_R')\Big)\frac{ \psi^B_-(\theta,n_L',n_R')}{R_{-1/2}^B(n_L',n_R')}\nonumber\\
{\bar \psi}_F (r,\theta, n_L,n_R)&=&\delta(n_L-n_L')\delta(n_R-n_R')\,\frac{{\bar \psi}^{B}_-(\theta,n_L',n_R')}{\bar R_{-1/2}^B(n_L',n_R')} \Big(\bar R_{1/2}(r,n_L',n_R')\Gamma^0 Z^{*-1}+\bar R_{-1/2}(r,n_L',n_R')\Big)\nonumber
\end{eqnarray}
where $n_L$ and $n_R$ are the momenta dual to the coordinates $t$ and $\phi$.
We insert $\psi$ and $\bar\psi$ into the boundary term
\begin{eqnarray}
\int d\theta\, \int dt\,  d\phi\, \sqrt{-g} \,\bar \psi \Gamma^r\psi\Big|_{r=r_B}&=&\int d\theta\, \sqrt{-g_B}\, \int dn_L' dn_R' \, \int dn_L'' dn_R''\,  \delta(n_L'-n_L'')\,  \delta(n_R'-n_R'')\times \nonumber\\
&&\times {\bar \psi}_F (r_B,\theta, n_L',n_R')\, \Gamma^r\, { \psi}_F (r_B, \theta,n_L'',n_R''),
\end{eqnarray}
where the determinant of the metric only depends on $\theta$ and the cutoff $r_B$
\begin{equation}
\sqrt{-g_B}=(2J\Gamma(\theta))^{3/2}\Lambda(\theta)\,  r_B.
\end{equation}
Using the explicit form of $\Gamma^r$ and the properties of the projection operator listed in appendix \ref{appB}, we can evaluate the integrand of the boundary term 
\begin{eqnarray}\label{bdr}
{\bar \psi}_F (r_B,\theta, n_L',n_R')&\Gamma^r&{ \psi}_F (r_B,\theta, n_L'',n_R'')=-\frac{r_B^2}{8J\Gamma}{\bar \psi}_+\Gamma^0\psi_+
+\frac{1}{2}{\bar \psi}_-\Gamma^0\psi_-\nonumber\\
&=&-\frac{r_B^2}{8J\Gamma}\frac{R^B_{1/2}\bar R^B_{1/2}}{R_{-1/2}^B
\bar R_{-1/2}^B}{\bar \psi}_-^B\Gamma^0 |Z^{-1}|^2\psi_-^B+\frac{1}{2}{\bar\psi}^B_-\Gamma^0 \psi_-^B,
\end{eqnarray}
where we have dropped the coordinate dependency on the rhs for simplicity of the notation. 
We would like to factor out the $\theta$-dependency of the boundary term. To do so notice that 
$\psi_-^{B}$  can be split into a two dimensional chiral spinor $\chi_0$ and a boundary spinor describing the theta 
dependence
\begin{equation}
\psi_{-}^B=\chi_{0}\otimes(S^+\oplus S^-).
\end{equation}
In the four dimensional representation space we know each spinor explicitly
\begin{equation}
\psi^{B}_-(\theta, n_R,n_L)=\underbrace{\delta(n_L-n_L')\delta(n_R-n_R')R_{-1/2}^B(n_L',n_R')}_{\chi_0(n_L',n_R')}
\left (
\underbrace{Z\left(
\begin{matrix}
0\\S_{-1/2}\\0\\0
\end{matrix}
\right)}_{S^+}+
\underbrace{Z\left(
\begin{matrix}
0\\0\\-S_{1/2}\\0
\end{matrix}
\right)}_{S^-}
\right ).
\end{equation}
\begin{equation}
\bar\psi^{B}_-(\theta, n_R,n_L)=\underbrace{\delta(n_L-n_L')\delta(n_R-n_R')\bar R_{-1/2}^B(n_L',n_R')}_{\bar \chi_0(n_L',n_R')}
\left (
\underbrace{Z\left(
\begin{matrix}
0\\S_{-1/2}\\0\\0
\end{matrix}
\right)}_{S^+}+
\underbrace{Z\left(
\begin{matrix}
0\\0\\-S_{1/2}\\0
\end{matrix}
\right)}_{S^-}
\right )^\dagger\, \Gamma^0.
\end{equation}

Inserting this into \eqref{bdr} the boundary term we notice that the theta dependence of the relevant contribution (the first term of the expression below) can be factored out
\begin{eqnarray}
&&\int d\theta \sqrt{-g_B}(|S_{1/2}|^2+|S_{-{1/2}}|^2) \int  dn_L' dn_R'\, \int \, dn_L'' dn_R''\, \bar \chi_0(n_L',n_R')\chi_0(n_L'',n_R''))\times\nonumber\\
&&\times\delta(n_L'-n_L'')\delta(n_R'-n_R'')\times \left(-\frac{r_B^2}{8J\Gamma}\frac{R^B_{1/2}(r_B, n_L'',n_R'')\bar R^B_{1/2}(r_B, n_L',n_R')}{R_{-1/2}^B(r_B, n_L'',n_R'')\bar R_{-1/2}^B(r_B, n_L',n_R')}\ +\frac{1}{4M^2(1+\cos ^2\theta)}\right)\, .\nonumber\\
\end{eqnarray}
The second term in this expression describes a contact term that can be ignored, so we evaluate the first 
term. To do so we expand $R_{1/2}$ and $R_{-1/2}$ around  $r_B$ 
using equations \eqref{R}-\eqref{beta} to evaluate individual contributions. We are left with

\begin{eqnarray}
&&\frac{{\delta}^2 S}{\delta\chi_0(n_L,n_R)\bar\delta\chi_0(n_L,n_R)}\sim r_B^3\,\frac{R^B_{1/2}( n_L,n_R)\bar R^B_{1/2}( n_L,n_R)}{R_{-1/2}^B(n_L,n_R)\bar R_{-1/2}^B( n_L,n_R)}\nonumber\\
&=&\frac{N_{1/2}}{N_{-1/2}}\, \frac{\bar N_{1/2}}{\bar N_{-1/2}}\,\left(\frac{A_{1/2}}{A_{-1/2}}\,\frac{\bar A_{1/2}}{\bar A_{-1/2}}r_B
 +\frac{B_{1/2}}{A_{-1/2}}\,\frac{\bar A_{1/2}}{\bar A_{-1/2}}T_R^{2\beta}r_B^{-2\beta+1}+\frac{A_{1/2}}{A_{-1/2}}\,\frac{\bar B_{1/2}}{\bar A_{-1/2}}T_R^{2\beta}r_B^{-2\beta+1}+{\cal O}(r^{-4\beta+1})\right)
\nonumber\\
&=&\frac{1}{M^2}r_B+\left(\mu+\frac{i\Lambda_\ell}{M}\right) \frac{\bar N_{1/2}\bar A_{1/2}} {\bar N_{-1/2}\bar A_{-1/2} }\, G_R(n_L,n_R)r_B^{-2\beta+1}+\left(\mu-i\frac{\Lambda_\ell}{M}\right) \frac{ N_{1/2} A_{1/2}} { N_{-1/2} A_{-1/2} }G^*_R(n_L,n_R)r_B^{-2\beta+1}\nonumber\\ &&\qquad\qquad\qquad\qquad\qquad\qquad\qquad\qquad\qquad\qquad\qquad\qquad\qquad\qquad\qquad\qquad+{\cal O}\left(r^{-4\beta+1}\right)
\end{eqnarray}
with
\begin{equation}\label{GR}
G_R(n_L,n_R)=-\frac{i}{\beta+in_L}\frac{\Gamma(-2\beta)}{\Gamma(2\beta)}\frac{\Gamma(\beta-in_L)}{\Gamma(-\beta-in_L)}\frac{\Gamma(\frac{1}{2}+\beta-in_R)}{\Gamma(\frac{1}{2}-\beta-in_R)}T_R^{2\beta}
\end{equation}
The first term above is obviously the contact term.
The second and third terms are complex conjugate to each other. Similarly as for the scalar two point function in Lorentzian AdS/CFT considered in \cite{Son:2002sd}, this means that the two point function is real, which is not what we want. 
The proposal of \cite{Son:2002sd} is to drop the complex conjugate solution. 
The $r_B$-factor here can be absorbed into $\chi_0$ by rescaling
\begin{equation}
\chi_0\rightarrow r_B^{\beta-1/2}\chi_0.
\end{equation}
Last, we factored out the ratio $\frac{N_{1/2}A_{1/2}}{N_{-1/2}A_{-1/2}}$ which is momentum dependent but not part of the two point function. A similar factor emerges in AdS/CFT calculations  \cite{Iqbal:2009fd}. 

The expression \eqref{GR} agrees with the proposal of \cite{Chen:2010ni}. Recall that the relation between $G_R$ and the absorption probability $\sigma$ is ${\rm{Im}}\, G_R\sim \sigma$. The Greens function we calculated precisely gives the absorption probability of \cite{Bredberg:2009pv}.\footnote{We thank Tom Hartman for pointing this out.}
\section{Fermionic Two-Point Function: CFT Result}\label{CFTcomp}

This section serves as a reminder for some basics on finite temperature conformal field theory.
We would like to write the finite temperature two-point function of a two dimensional CFT in momentum space and compare with the result of the previous section. 
We start with the more familiar zero temperature correlation function in coordinate space.
The zero temperature two-point function of a conformal field theory operator with conformal weights 
\begin{equation}
h_L=\frac{1}{2}\left(\Delta-\frac{1}{2}\right)\ , \qquad h_R=\frac{1}{2}\left(\Delta+\frac{1}{2}\right)\ ,
\end{equation}
takes (up to a constant) the following form in coordinate space
\begin{equation}\label{retgreens}
\langle {\cal O}(\vec x)\bar {\cal O}(\vec y)\rangle 
\sim\frac{\gamma^i(x^i-y^i)}{|\vec x-\vec y|^{(2\Delta+1)}} \ .
\end{equation}
More explicitly we can use the following representation of the gamma matrices
\begin{equation}\label{gammas}
\gamma^0=\left(
\begin{array}{cc}
0&-1\\1&0
\end{array}
\right)\ ,
\qquad
\gamma^1=\left(
\begin{array}{cc}
0&1\\1&0
\end{array}
\right) \ , 
\end{equation}
and the coordinates 
\begin{eqnarray}
t^+_1&=&x^0+x^1 \ , \qquad t^-_1=x^0-x^1\ ,\nonumber\\
t^+_2&=&y^0+ y^1 \ , \qquad t^-_2=y^0-y^1 \ ,\nonumber
\end{eqnarray}
to rewrite (\ref{retgreens}) as 
\begin{equation}
\langle {\bar \cal O}(t_1^+,t_1^-) {\cal O}(t_2^+,t_2^-)\rangle \sim \gamma^0
\left(
\begin{array}{cc}
\frac{1} { 
(t^+_{12})^{2h_R-1} (t^-_{12})^{2h_L+1}
}&0\\
0&\frac{1}{{(t^+_{12})}^{2h_R}{(t^-_{12})}^{2h_L}}
\end{array}
\right) \ ,
\end{equation}
where we have introduced $t_{12}^+=t_1^+-t_2^+$ and similarly for $t_{12}^-$.
The finite temperature correlation function is obtained by mapping the above result  to a torus with circumferences $1/T_L$ and $1/ T_R$
\begin{equation}\label{resexpr}
\langle {\cal O}(t_1^+,t_1^-)\bar {\cal O}(t_2^+,t_2^-)\rangle
\sim\gamma^0\left(
\begin{array}{cc}
\left(\frac{\pi T_R}{\sinh(\pi T_R t^+_{12})}\right)^{2h_R-1} \left(\frac{\pi T_L}{\sinh(\pi T_L t^-_{12})}\right)^{2h_L+1} &0\\
0&\left(\frac{\pi T_R}{\sinh(\pi T_R t^+_{12})}\right)^{2h_R} \left(\frac{\pi T_L}{\sinh(\pi T_L t^-_{12})}\right)^{2h_L}
\end{array}
\right).
\end{equation}

The formula (\ref{resexpr}) is the two-point function $\langle {\cal O}\bar {\cal O}\rangle$ for a non-chiral spinor operator ${\cal O}$. The AdS/CFT correspondence gives a correlator only between chiral/antichiral  parts of the operator ${\cal O}$.
\begin{eqnarray}\label{chiral}
{\cal O}^{\pm}=\frac{1}{2}\left(1\pm\gamma^0\gamma^1\right){\cal O} \ ,\qquad \bar{\cal O}^{\pm}=\bar {\cal O}\frac{1}{2}\left(1\mp\gamma^0\gamma^1\right) \ .
\end{eqnarray}
Inserting (\ref{chiral}) into $\langle {\cal O}\bar {\cal O}\rangle$ with ${\cal O}=\left(\begin{array}{c}{\cal O}_1\\{\cal O}_2\end{array}\right)$ we see that the non-zero elements of (\ref{resexpr}) can be identified with 
\begin{equation}
\langle {\cal O}^+\bar {\cal O}^+\rangle=\gamma^0\left(
\begin{array}{cc}
\langle{\cal O}_2 {\cal O}_2\rangle & 0\\
0&0
\end{array}
\right) \ ,\qquad 
\langle {\cal O}^-\bar {\cal O}^-\rangle=\gamma^0\left(
\begin{array}{cc}
0&0\\ 0&
\langle{\cal O}_1 {\cal O}_1\rangle
\end{array}
\right)\, .
\end{equation}

After analytic continuation $t^\pm\rightarrow it^\pm$, we Fourier transform the two-point function assuming only integer frequencies $\omega_E=2\pi k T$ by using 
\begin{equation}
\int_{0}^{1/T}dt e^{i\omega_E t}\left(\frac{\pi T}{\sin(\pi Tt)}\right)^{2h}= \frac{(\pi T)^{2h-1}2^{2h}e^{i\omega_E/2T}\Gamma(1-2h)}{\Gamma\left(1-h+\frac{\omega_E}{2\pi T}\right)\Gamma\left(1-h-\frac{\omega_E}{2\pi T}\right)}\ , .
\end{equation}

 Once we identify 
$k_L=-in_L , k_R=-in_R$ and $h_L=\beta ,  h_R=\beta+\frac{1}{2}, T_L=\frac{1}{2\pi}, T_R=T_R$ the two-point function $\langle {\cal O}^-{\bar{\cal O}}^-\rangle$ on the CFT side 
becomes\footnote{Here we have absorbed the $m\Omega_R$ appearing in eq. (5.13) of \cite{Bredberg:2009pv} into our definition of $n_R$.}
\begin{equation}
\langle {{\cal O}}^-{\bar{\cal O}}^-\rangle\sim T_R^{2\beta}\frac{1}{\beta+in_L}\frac{\Gamma(-2\beta)\Gamma(\beta-in_L)\Gamma(\frac{1}{2}+\beta-in_R)}{\Gamma(2\beta)\Gamma(-\beta-in_L)\Gamma(\frac{1}{2}-\beta-in_R)}\, .
\end{equation}
This matches the expression computed on the bulk side. 

\section{Conclusions}
In this note we have calculated finite temperature two point correlations functions for fermionic fields in Kerr/CFT using the variational principle.
Fermionic fields are particularly interesting because their correlation functions describe semi-realistic physical observables, such as the spectral function. 

To perform this calculation we have followed an approach well known for AdS/CFT. 
After analyzing the variational principle we have seen that a boundary term needs to be added to the Dirac action for the variational principle to be satisfied. This boundary term is responsible for generating non-trivial fermion correlation functions.
Kerr/CFT is a duality in which a four-dimensional bulk geometry is dual to a two-dimensional conformal field theory. The fact that the conformal field theory lives in two dimensions less than the original bulk theory may sound at first surprising because from AdS/CFT we are used to the fact that the conformal field theory lives in one dimension less than the bulk rather than two. 
Fermions allow us to very nicely understand this aspect of Kerr/CFT because fermions, as opposed to scalars,
are very sensitive to the number of space-time dimensions they live in.
The boundary of the near-NHEK geometry is a three-dimensional theory described by the coordinates $t$,$\phi$ and $\theta$, while the radial coordinate approaches a large but finite cutoff $r_B$. 
Performing the calculation of the two point-function for two spinors  living on the 3D boundary, we have seen that the theta dependence of the correlation function factors out. Therefore, the fermion correlation function effectively becomes that of a two dimensional relativistic conformal field theory. 

Our calculation was performed in Lorentzian signature rather than with an analytic continuation to Euclidean signature. We are not aware of a sensible Euclidean analytic continuation of the near-NHEK metric. For this reason we needed to impose some additional constraints on the two-point function that are well known from Lorentzian approaches to AdS/CFT \cite{Son:2002sd}.

An interesting observation is that the gravitational action functional needed the inclusion of a boundary term that breaks Lorentz invariance and one may wonder if the boundary conformal field theory could be a non-relativistic theory once corrections to the leading terms are included. This would be similar in spirit to the recent discussion appearing in \cite{Laia:2011zn} in the context of AdS/CFT. Here a bulk theory in $AdS_4$ space-time is supplied with boundary conditions on the spinor field that break Lorentz invariance and it is argued that the dual conformal field theory is non-relativistic. Some recent discussion on the connection between Kerr/CFT and non-relativistic conformal field theories has recently appeared in \cite{ElShowk:2011cm}.
It would be interesting to explore this connection in more detail.

Finally,  it would be interesting to extend our calculation to the Kerr-Newman geometry, as well as to other correlation functions involving e.g fermions and gauge fields. We hope to report on this in the future. 

\section*{Acknowledgments}
We benefited from discussions with Katrin Becker, David Chow, Sera Cremonini, Umut Gursoy, Chris Pope,  Daniel Robbins, Jan Troost as well as the correspondence with Aaron Amsel, Geoffrey Compere, Monica Guica, Tom Hartman, Gary Horowitz, Wei Song and Andrew Strominger.  We would like to thank Tom Hartman and Andrew Strominger for comments on the manuscript. This work was supported by NSF under PHY-0505757, DMS-0854930 and the University of Texas A\&M.

\appendix

\section{Near-NHEK Geometry}\label{appA}

The Kerr/CFT correspondence relates the near horizon geometry of a near extreme Kerr black hole (near NHEK) to a two-dimensional conformal field theory with central charges $c_L=c_R=12J/\hbar$. 
The near-NHEK metric is constructed by taking a special limit of the Kerr metric. Let us summarize the main steps of this construction, as they are needed for the calculation of the fermionic two-point function. 
The geometry of the Kerr black hole is described by the metric
\begin{equation}\label{kerrmetric}
ds^2=-\frac{\Delta}{\hat \rho^2}\left(d\hat t-a\sin^2\theta d\hat\phi\right)^2+\frac{\sin^2\theta}{\hat \rho^2}\left(\left(\hat r^2+a^2\right)d\hat\phi-a d\hat t\right)^2+\frac{\hat \rho^2}{\Delta}d\hat r^2+\hat\rho^2d\theta^2,
\end{equation} 
with $\Delta=\hat r^2-2M\hat r +a^2$, $\hat\rho^2=\hat r^2+a^2\cos^2\theta$. In general, there are two horizons at
\begin{equation}
r_\pm=M\pm\sqrt{M^2-a^2},
\end{equation}
where $a$ is the proportionality factor between the angular momentum and the mass $J=aM$. The Hawking temperature and the angular velocity of the horizon are
\begin{equation}
T_H=\frac{r_+-r_-}{8\pi Mr_+}=\frac{\tau_H}{8\pi M}\ , \qquad\Omega_H=\frac{a}{2Mr_+}  \ .
\end{equation}
The near horizon limit of the near extremal Kerr black hole can be defined by taking the limit $T_H\rightarrow 0, \ \hat r\rightarrow r_+$ with the dimensionless near-horizon temperature $T_R=\frac{2MT_H}{\lambda}$ fixed when $\lambda\rightarrow 0$. Following \cite{Hartman:2009nz} the metric of the near-NHEK space-time is obtained by performing the expansions
\begin{equation}
r_+=M+\lambda M 2\pi T_R+{\cal O} (\lambda^2) \ , \qquad a=M-2M (\lambda\pi T_R)^2 +{\cal O}(\lambda^3) \ ,
\end{equation}
coordinate redefinitions
\begin{equation}\label{coordredef}
t=\lambda \frac{\hat t}{2M} \ , \qquad r=\frac{\hat r-r_+}{\lambda r_+} \ , \qquad \phi=\hat\phi-\frac{\hat t}{2M},
\end{equation}
and taking limit $\lambda\ll 1$ while keeping $T_R$ fixed. The near-NHEK metric is then
\begin{equation}\label{nearNHEK}
ds^2=2J\Gamma\Big(-r(r+4\pi T_R)dt^2+\frac{dr^2}{r(r+4\pi T_R)}+d\theta^2+\Lambda^2\left(d\phi+(r+2\pi T_R)dt\right)^2\Big) \ ,
\end{equation}
where
\begin{equation*}
\Gamma(\theta)=\frac{1+\cos^2\theta}{2} \ , \qquad \Lambda(\theta)=\frac{2\sin\theta}{1+\cos^2\theta} \ , \qquad\phi\sim\phi+2\pi,\ 0\le\theta\le\pi\ .
\end{equation*}

The appearance of $\lambda$ in (\ref{coordredef}) may look confusing. The range of the near-NHEK space is parametrized by the radial coordinate $r$, where it takes values $0< \lambda r\ll1$. Notice that since $\lambda\ll1$, the position of the near-NHEK boundary is at some large but still finite value of $r$.


\section{Notations and Conventions}\label{appB}

\begin{itemize}
\item  We define
$\slashed D=\Gamma^MD_M$, 
$D_M=\partial_M+\frac{1}{4}\omega_{ab\,M}\Gamma^{ab}$ with $\omega$ being the bulk spin connection, $\Gamma^{ab}={1\over 2}[\Gamma^a, \Gamma^b]$, while the conjugate spinor is defined as $\bar \psi={\psi}^{\dagger}{\Gamma}^0$. Capital indices denote bulk space-time indices and $a,b$ denote bulk tangent indices.

\item The flat gamma matrices are
\begin{eqnarray}
\Gamma^0&=&\left(
\begin{matrix}
0&0&1&0\\
0&0&0&1\\
1&0&0&0\\
0&1&0&0
\end{matrix}
\right) , \ \
\Gamma^1=\left(
\begin{matrix}
0&0&0&-1\\
0&0&-1&0\\
0&1&0&0\\
1&0&0&0
\end{matrix}\right), \ \ 
\nonumber\\
\Gamma^2&=&\left(
\begin{matrix}
0&0&0&i\\
0&0&-i&0\\
0&-i&0&0\\
i&0&0&0
\end{matrix}
\right) , \ \ 
\Gamma^3=\left(
\begin{matrix}
0&0&-1&0\\
0&0&0&1\\
1&0&0&0\\
0&1&0&0
\end{matrix}\right)\nonumber\\
\end{eqnarray}

\item The curved gamma for the near NHEK geometry  are given in \cite{Hartman:2009nz} 
\begin{equation}
\Gamma^\mu=\sqrt{2}\left(
\begin{matrix}
0&\sigma^\mu_{AB'}\\ \bar \sigma^{\mu A' B} &0
\end{matrix}
\right)\ , \qquad
\sigma_{AB'}^\mu=\left(
\begin{matrix}
l^\mu &m^\mu\\
\bar m^\mu & n^\nu
\end{matrix}
\right)\, ,
\end{equation}
where the Newman-Penrose tetrad for the near-NHEK was worked out in \cite{Chen:2010ni}
\begin{eqnarray}
l^\mu&=&\frac{1}{r(r+4\pi T_R)}(1,r(r+4\pi T_R),0,-(r+2\pi T_R))\ ,\nonumber\\
n^\mu &=&\frac{1}{4 J\Gamma (\theta)}(1,-r(r+4\pi T_R),0,-(r+2\pi T_R))\ ,\nonumber\\
m^\mu&=&\frac{1}{2\sqrt{J\Gamma(\theta)}}(0,0,1,i\Lambda^{-1}(\theta)) \ .
\end{eqnarray}

For the near-NHEK geometry $\Gamma^r$ is given by
\begin{eqnarray}
\Gamma^r&=&\left(
\begin{matrix}
0&0&1&0\\
0&0&0&-\frac{r(r+4\pi T_R)}{4J\Gamma}\\
-\frac{r(r+4\pi T_R)}{4J\Gamma}&0&0&0\\
0&1&0&0
\end{matrix}
\right)\\
&=&-\frac{r(r+4\pi T_R)}{8J\Gamma(\theta)}(\Gamma^0+\Gamma^3)+\frac{1}{2}(\Gamma^0-\Gamma^3)
\end{eqnarray}

\end{itemize}

\bibliographystyle{utphys}
\bibliography{kerrcftspinors}

\providecommand{\href}[2]{#2}\begingroup\raggedright\begin{thebibliography}{10}

\bibitem{Guica:2008mu}
M.~Guica, T.~Hartman, W.~Song, and A.~Strominger, ``{The Kerr/CFT
  Correspondence},'' \href{http://dx.doi.org/10.1103/PhysRevD.80.124008}{{\em
  Phys.Rev.} {\bfseries D80} (2009) 124008},
\href{http://arxiv.org/abs/0809.4266}{{\ttfamily arXiv:0809.4266 [hep-th]}}.

\bibitem{Bredberg:2011hp}
I.~Bredberg, C.~Keeler, V.~Lysov, and A.~Strominger, ``{Cargese Lectures on the
  Kerr/CFT Correspondence},''
  \href{http://dx.doi.org/10.1016/j.nuclphysbps.2011.04.155}{{\em
  Nucl.Phys.Proc.Suppl.} {\bfseries 216} (2011) 194--210},
\href{http://arxiv.org/abs/1103.2355}{{\ttfamily arXiv:1103.2355 [hep-th]}}.

\bibitem{Hartman:2009nz}
T.~Hartman, W.~Song, and A.~Strominger, ``{Holographic Derivation of
  Kerr-Newman Scattering Amplitudes for General Charge and Spin},''
  \href{http://dx.doi.org/10.1007/JHEP03(2010)118}{{\em JHEP} {\bfseries 1003}
  (2010) 118},
\href{http://arxiv.org/abs/0908.3909}{{\ttfamily arXiv:0908.3909 [hep-th]}}.

\bibitem{Chen:2010ni}
B.~Chen and C.-S. Chu, ``{Real-Time Correlators in Kerr/CFT Correspondence},''
  \href{http://dx.doi.org/10.1007/JHEP05(2010)004}{{\em JHEP} {\bfseries 1005}
  (2010) 004},
\href{http://arxiv.org/abs/1001.3208}{{\ttfamily arXiv:1001.3208 [hep-th]}}.

\bibitem{Henningson:1998cd}
M.~Henningson and K.~Sfetsos, ``{Spinors and the AdS / CFT correspondence},''
  \href{http://dx.doi.org/10.1016/S0370-2693(98)00559-0}{{\em Phys.Lett.}
  {\bfseries B431} (1998) 63--68},
\href{http://arxiv.org/abs/hep-th/9803251}{{\ttfamily arXiv:hep-th/9803251
  [hep-th]}}.

\bibitem{Mueck:1998iz}
W.~Mueck and K.~Viswanathan, ``{Conformal field theory correlators from
  classical field theory on anti-de Sitter space. 2. Vector and spinor
  fields},'' \href{http://dx.doi.org/10.1103/PhysRevD.58.106006}{{\em
  Phys.Rev.} {\bfseries D58} (1998) 106006},
\href{http://arxiv.org/abs/hep-th/9805145}{{\ttfamily arXiv:hep-th/9805145
  [hep-th]}}.

\bibitem{Henneaux:1998ch}
M.~Henneaux, ``{Boundary terms in the AdS / CFT correspondence for spinor
  fields},''
\href{http://arxiv.org/abs/hep-th/9902137}{{\ttfamily arXiv:hep-th/9902137
  [hep-th]}}.

\bibitem{Iqbal:2009fd}
N.~Iqbal and H.~Liu, ``{Real-time response in AdS/CFT with application to
  spinors},'' \href{http://dx.doi.org/10.1002/prop.200900057}{{\em
  Fortsch.Phys.} {\bfseries 57} (2009) 367--384},
\href{http://arxiv.org/abs/0903.2596}{{\ttfamily arXiv:0903.2596 [hep-th]}}.

\bibitem{Laia:2011zn}
J.~N. Laia and D.~Tong, ``{A Holographic Flat Band},''
  \href{http://dx.doi.org/10.1007/JHEP11(2011)125}{{\em JHEP} {\bfseries 1111}
  (2011) 125}, \href{http://arxiv.org/abs/1108.1381}{{\ttfamily arXiv:1108.1381
  [hep-th]}}.
21 Pages, 15 Figures. v2: Reference and Acknowledgement added.

\bibitem{Marolf:2004fy}
D.~Marolf, ``{States and boundary terms: Subtleties of Lorentzian AdS / CFT},''
  \href{http://dx.doi.org/10.1088/1126-6708/2005/05/042}{{\em JHEP} {\bfseries
  0505} (2005) 042},
\href{http://arxiv.org/abs/hep-th/0412032}{{\ttfamily arXiv:hep-th/0412032
  [hep-th]}}.

\bibitem{Son:2002sd}
D.~T. Son and A.~O. Starinets, ``{Minkowski space correlators in AdS / CFT
  correspondence: Recipe and applications},'' {\em JHEP} {\bfseries 0209}
  (2002) 042,
\href{http://arxiv.org/abs/hep-th/0205051}{{\ttfamily arXiv:hep-th/0205051
  [hep-th]}}.

\bibitem{Chandrasekhar:1976ap}
S.~Chandrasekhar, ``{The Solution of Dirac's Equation in Kerr Geometry},''
{\em Proc.Roy.Soc.Lond.} {\bfseries A349} (1976) 571--575.

\bibitem{ElShowk:2011cm}
S.~El-Showk and M.~Guica, ``{Kerr/CFT, dipole theories and nonrelativistic
  CFTs},''
\href{http://arxiv.org/abs/1108.6091}{{\ttfamily arXiv:1108.6091 [hep-th]}}.

\bibitem{Martellini:1977qf}
M.~Martellini and A.~Treves, ``{Absence of Superradiance of a Dirac Field in a
  Kerr Background},''
\href{http://dx.doi.org/10.1103/PhysRevD.15.3060}{{\em Phys.Rev.} {\bfseries
  D15} (1977) 3060--3061}.

\bibitem{Iyer:1978du}
B.~R. Iyer and A.~Kumar, ``{Note on the Absence of Massive Fermion
  Superradiance from a Kerr Black Hole},''
\href{http://dx.doi.org/10.1103/PhysRevD.18.4799}{{\em Phys.Rev.} {\bfseries
  D18} (1978) 4799--4801}.

\bibitem{Bredberg:2009pv}
I.~Bredberg, T.~Hartman, W.~Song, and A.~Strominger, ``{Black Hole
  Superradiance From Kerr/CFT},''
  \href{http://dx.doi.org/10.1007/JHEP04(2010)019}{{\em JHEP} {\bfseries 1004}
  (2010) 019},
\href{http://arxiv.org/abs/0907.3477}{{\ttfamily arXiv:0907.3477 [hep-th]}}.

\end{thebibliography}\endgroup

\end{document}